# Reaction-diffusion pathways for a programmable nanoscale texture of diamond-SiC composite

Vladimir Ya. Shevchenko[a,b], Aleksei I. Makogon [a,c,*], Maxim M. Sychov [a,c], Michael Nosonovsky[d,e,*], Ekaterina V. Skorb [d]

[a] Institute of Silicate Chemistry, Russian Academy of Sciences, St. Petersburg, 199034 Russia
[b] NRC Kurchatov Institute - CRISM Prometey, 49 Shpalernaya str., St. Petersburg, 191015, Russia
[c] St. Petersburg State Institute of Technology St. Petersburg, 190013, Russia
[d] ITMO University, 9 Lomonosov St., St. Petersburg, 191002, Russia
[e] University of Wisconsin-Milwaukee, 3200 N Cramer St., Milwaukee, WI 53210, USA

[*]Corresponding author. Tel. +7-911-9825560, makogon422833@gmail.com; Tel.  +1-414-2292816, nosonovs@uwm.edu

**Abstract**

The diamond–SiC composite has a low density and the highest possible speed of sound among existing materials except for the diamond. The composite is synthesized by a complex exothermic chemical reaction between diamond powder and liquid Si. This makes it an ideal material for protection against impact loading. Experiments show that a system of patterns is formed at the diamond-SiC interface. Modeling of reaction-diffusion processes of composite synthesis proves a formation of ceramic materials with a regular (periodic) interconnected microstructure in a given system. Composite material with interconnected structures at the interface has very high mechanical properties and resistance to impact since its fractioning is intercrystallite.

**Introduction**

High-strength materials that are able to resist impact are needed for various practical applications including armor. Such materials usually have high hardness and high elastic moduli, facilitating also high speeds of elastic waves (sound), so that resulting shock waves quickly dissipate impact energy. Based on theoretical considerations, there is a limit on the maximum possible speed of elastic waves in the medium sound which is estimated as [1]

$$V_s = c\alpha \sqrt{\frac{m_e}{2m_p}} \approx 36.12 \cdot 10^3 \text{ m/s} \qquad (1)$$

This is because the speed of longitudinal elastic waves depends on the elastic bulk and shear moduli ($K$ and $\mu$), and on the density $\rho$ of the material, $V_s = \sqrt{\frac{K+\frac{4}{3}\mu}{\rho}} \approx \sqrt{\frac{E}{m}}$, where $E$ is the bonding energy and $m$ is the mass of the atom or molecule. The bonding energy can be further reduced through the Rydberg energy to fundamental constants yielding Eq. 1.

However, it is difficult to reach the theoretical limit of the speed of sound in solid materials. This is because these materials usually have defects, such as grains, heterogeneities, and dislocations, which significantly reduce the speed of elastic waves. Thus, diamond has the highest practically observed speed of sound has of about 18,000 m/s.

While diamond provides very high stiffness, it cannot be used as a monolithic material, because it is impossible to manufacture products of complex shape and significant volume. To overcome this obstacle, composite materials can be used with diamond is embedded in a ceramic matrix [2]. Of particular interest is diamond - silicon carbide (SiC) composites, because SiC possesses very high elastic modulus and other mechanical properties, as well as a structural and chemical affinity to diamond, i.e., it forms a coherent interface with it [3-4]. Such composite was synthesized by Shevchenko et al. [5] under the commercial name *Ideal*.

Table 1 shows the physical and mechanical properties of the material used for ceramic armors. The high speed of propagation of sound waves is important for several reasons:

1) the impact energy dissipates faster over a larger area of the ceramic;

2) the volume of the striker's damage increases.

The greater the difference in the velocity of propagation of shock waves in the material of the impactor and the ceramic material, the more the impactor will be destroyed and stopped. In steel, the shock waves propagate at a speed of 4.9 km / s, while in the SiC at 9-10 km/s, in the boron carbide at 12-13 km/s, and in the diamond at SiC composite material *Ideal* at 15 km/s (**Table 1**).

In the work of Shevchenko et al. [5], the parameter characterizing the armor resistance of a material based on the speed of sound is discussed:

$$K = \frac{l}{l_{steel}} = \sqrt[3]{\frac{\rho_{steel} \cdot V_{steel}^2}{\rho_{cer} \cdot V_{cer}^2}} \qquad (2)$$

where $l_{steel}$ is the initial length of the steel impactor used to test the ceramics for armor resistance, $l$ is the length of the impactor after interaction with the obstacle, $V$ is the longitudinal speed of sound, $\rho_{steel}$ is the density of steel, and $\rho_{cer}$ is the density of ceramic.

Based on this approach, diamond is the best armor material, while the second closest to it in properties is the material *Ideal* developed by the authors [3]. *Ideal* is a diamond–SiC composite synthesized during the process of reaction-diffusion transformations.

**Table 1. Properties of ceramic impact-resistant materials**

| Material | Density, $\rho$, kg/m$^3$ | Sound speed, $V$, m/s | Elastic modulus, $E$, GPa |
|---|---|---|---|
| Sintered corundum (Al$_2$O$_3$) | 3750 | 9800 | 375 |
| Reaction sintered Silicon Carbide (SiSiC) | 3100 | 10300 | 329 |
| Liquid Phase-Sintered Silicon Carbide (LPSSiC) | 3250 | 10500 | 358 |
| Reaction-sintered Boron Carbide (RSB$_4$C) | 2550 | 13000 | 383 |
| Hot-pressed Boron Carbide (HPB$_4$C) | 2750 | 11800 | 431 |
| WC + 6% Co alloy | 15000 | 6500 | 633 |
| Sintered Titanium Diboride (TiB$_2$) | 4500 | 11000 | 540 |
| Diamond-silicon carbide composite (*Ideal*) | 3350 | 15000 | 754 |

In this paper, we will present the process of synthesis of the diamond–silicon carbide (SiC) composites and will show how this process results in the formation of self-organized patterns called "Turing structures" depending on concentrations of reactants at interfaces. Theoretical models and experimental confirmation will be provided.

There are many studies describing the formation of regular microstructures. However, for solid-phase systems, such processes have not been experimentally studied until recently. Shoji [6] was the first to suggest that Turing patterns in three-dimensional space show a connection between reaction-diffusion equations and surface geometry and topology. This is the basis of our idea of using reaction-diffusion processes in solid-phase systems for the formation of materials with a regular (periodic) interconnected microstructure, including those with the topology of three times periodic surfaces of minimum energy [7-8].

**Experimental**

The synthesis of the diamond–SiC involved several stages: (i) forming samples from a mixture of diamond powders, (ii) drying the samples, (iii) impregnating them with liquid Si, and (iv) cleaning the surface. The raw samples were formed by pressing the mixture in a metal mold with subsequent mechanical processing. The resulting porous specimens were siliconized by impregnating them with liquid Si. At this stage, the diamond grains were covered by the synthesized SiC, and the structure of the non-porous composite material was formed. After the synthesis of the SiC matrix, the surface of the materials was cleaned from technological contaminants by sandblasting. Details of this process are provided in the papers [7-13].

In order to examine the formed composites, and in particular, the silicon carbide-diamond interface, with a scanning double-beam transmission electron microscope (TEM), a thin slice of the surface layer with a size of 15×10×2 μm was prepared. Then the manipulator needle coated with a platinum layer was brought to the specimen. After that, the specimen was cut off from a massive sample by an ion beam, transferred by a manipulator to a special semi-grid TEM holder, and coated with a layer of platinum. Then, the manipulator needle was cut off from the specimen with an ion beam, and the specimen containing the interface was fixed in the semi-grid and thinned by the ion beam to 50-100 nm thickness. This thickness makes the cross-section transparent to electrons and allows it to be examined with FEI Tecnai G2 TEM. **Fig. 1** shows enlarged images of the interface, on which the morphological features are clearly manifested. These features form a partially ordered nanocomposite structure of the diamond – silicon carbide phases. No signs of interfacial separation can be observed. The identification of the observed phases was carried out according to the data of microdiffraction patterns obtained with TEM.

**Results and discussion**

*Experimental results.* The TEM analysis showed that there is a region at the diamond–SiC interface, where a mixture of the SiC phases (dark areas) with the diamond phase (light areas) is located, and the size of this region is up to 200 nm. The diffraction patterns were obtained using the single reflection method. The SiC phase was identified as the "3C" cubic modification. A comparison of diffraction patterns obtained from dispersed crystallites at the diamond–SiC interface allowed us to conclude that the dispersed crystallites observed at the interface have a high affinity with the diamond lattice. This, along with the interpenetrating structure of the phases, results in a high mechanical modulus of the material, as well as impact resistance under dynamic loading.

TEM images also showed the formation of sections with periodic structures at the interface, both perpendicular to the interface plane (**Fig. 1a, b**). It was concluded that these interpenetrating structures at the phase interface provide outstanding mechanical properties of the composite due to the high adhesion strength of diamond grains and silicon carbide matrix, **Fig. 1c**.

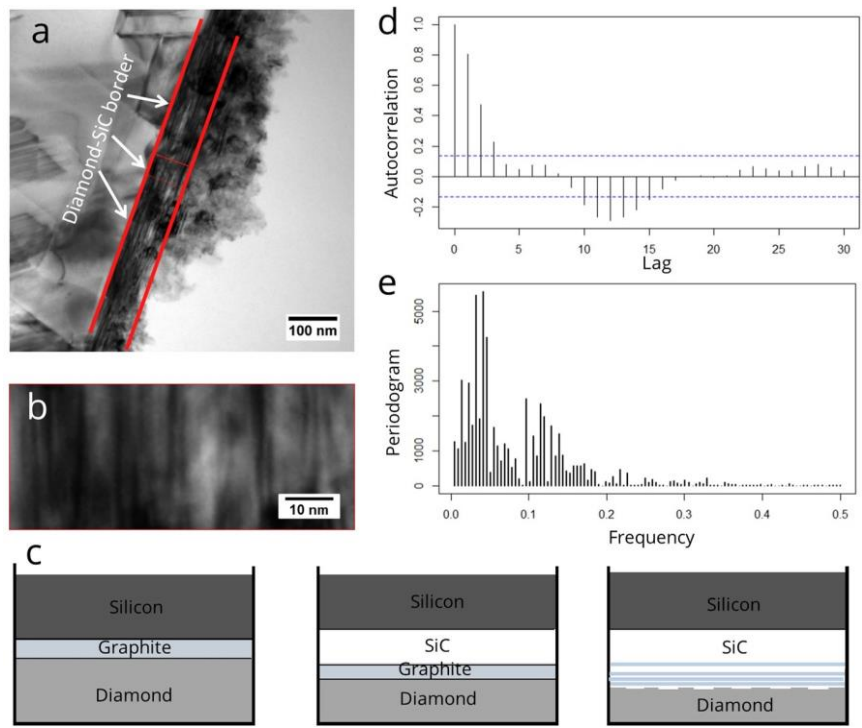

**Fig 1**. (a) Silicon carbide (left) - diamond (right) interface with clusters of small crystals, (b) considered area of the interface, (c) the scheme of the formation of a diamond - silicon carbide composite, (d) autocorrelation of the intensity of the color of the considered area along the red line, (e) the periodogram of the intensity of patterns.

The interpenetrating structure also results in phase alternation at the plane parallel to the diamond-silicon carbide interface. It is known that the lattice constants of diamond and silicon carbide differ significantly, which leads to the formation of internal stresses. Phase alternation/rotation reduces this effect.

Let us now consider the interface in more detail. **Fig. 1a** shows that stripes are formed on the surface of diamonds with a certain periodicity. The following method was used to analyze the periodicity. The analyzed area was considered separately (**Fig. 1b**), and an area with a size of 212×100 px (100×50 nm) was selected. Using the developed *Python* script, the value of the intensity of the white color (the sum of the values of the brightness of the red, blue, and green components) was calculated at each point of the line marked in **Fig. 1b**. Further analysis was carried out using scripts written in the *R* programming language. For the analysis of the periodicity, the autocorrelation function was used (**Fig. 1d**). If the original function is strictly periodic, then the plot of the autocorrelation function will also have a strictly periodic profile. Thus, from this plot, one can judge the periodicity of the original function, and, consequently, its frequency characteristics (**Fig. 1e**). The observed periodicity is nanoscale, on the scale order between nanometers and dozens of nanometers.

From the analysis of the above, one can conclude that the interface is periodic and the periodicity can be described by a function with two wavelengths. As mentioned above, in the direction parallel to the diamond – SiC interface (**Fig. 1a**), a region with periodicity is also observed (dark and light areas); however, due to its small size, it is difficult to numerically estimate the periodicity parameters for it. It is seen that these areas do not form a uniform layer at the interface, but propagate as semi-ovals from certain centers on the surface (the direction of formation in the example on the right is shown by arrows). During the synthesis of a diamond – SiC composite, the Si melt interacts primarily with graphite on the diamond surface. The graphitization process begins at defective areas of the diamond surface, gradually forming the so-called graphitization. It can be concluded that the process of forming composite material in the form of an interpenetrating structure begins at the defective areas of the diamond surface. Additional SEM images of the top view and TEM images of the side view of formed structures are shown in **Fig. 2**.

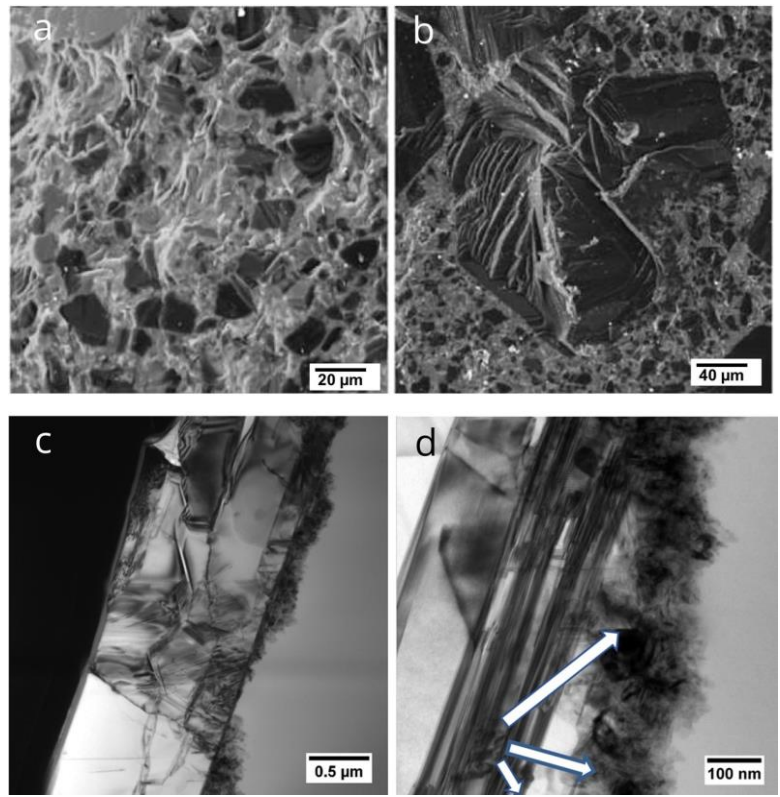

**Fig 2**. SEM images of top view of formed structures with visible darker points of diamonds of sized (a) from 10 µm (b) to more than 200 µm. (c)TEN of interphase boundary silicon carbide and (d) diamond with foci-accumulations of small crystals.

It is known that the formation of a structure with alternating phases (with the formation of patterns) is often caused by the reaction-diffusion Turing mechanism [14].

*Reaction-diffusion model of Diamond-SiC composite synthesis.* In 1952, Alan Turing suggested a system of partial differential equations (PDEs), which can describe the evolution of non-linear chemical systems yielding periodic spatial patterns, which was called the reaction-diffusion system [14]. Mathematically, reaction-diffusion systems are represented by semi-linear parabolic partial differential equations with the general matrix form

$$\partial_t \mathbf{a} = \mathbf{D}\nabla^2 \mathbf{a} + \mathbf{R}(\mathbf{a}), \qquad (3)$$

where $\mathbf{a}(x,y,z)$ is a function of spatial coordinates (typically $\mathbf{a}$ represents a column of concentration variable), $\mathbf{D}$ is the matrix of diffusion coefficients, $\mathbf{R}(\mathbf{a})$ represents functions, describing local reactions.

The solutions of the reaction-diffusion equations demonstrate a wide range of behaviors, including the formation of traveling waves and wave-like phenomena as well as other self-organized patterns like stripes, hexagons, or more intricate structures like dissipative solitons. Such patterns are referred to as Turing structures [15]. Turing structures can form when a stable steady-state system becomes sensitive to small heterogeneous perturbations due to diffusion rate imbalances. Thus, Tan et al. applied interfacial polymerization to grow polyamide membranes using the reactions at the interface between oil and water layers [16]. Turing patterns appear also in closed two-layer gel reactors [17], they are used for energy storage [18]. Of particular interest are 3D Turing patterns [19].

The formation of an interpenetrating composite in the diamond–SiC system for the 1D case can be considered in accordance with the approach described in [7-8]. To describe the synthesis of a diamond–SiC composite, the following model of chemical transformations was proposed.

When a diamond is heated, graphitization occurs on its surface. The presence of oxygen sorbed by the surface accelerates this process by a catalytic mechanism. This process can be described by the reactions

$$C_{diamond} + O_2 = C^*O_2 \qquad (4)$$

$$C^*O_2 = C_{graphite} + O_2 \qquad (5)$$

where $C_{diamond}$ and $C_{graphite}$ are carbon in the form of diamond and graphite, $C^*O_2$ is an activated complex.

Upon further heating above the melting temperature $T_{melt}=1414$ °C, silicon melts and wets the surface of the diamond particles. Consequently, the graphite formed on the diamond surface interacts with molten silicon, forming cubic silicon carbide:

$$C_{graphite} + S_{liq} = SiC_{cubic} \qquad (6)$$

The model also takes into account that silicon carbide can interact with oxygen, thereby reducing surface graphitization, while, from thermodynamic considerations, it follows that mainly CO gas is formed:

$$SiC_{cube} + 1.5O_2 = SiO_2 + CO \tag{7}$$

Carbon monoxide dissociates to form carbon, which is capable of interacting with silicon:

$$2CO = C_{graphite} + CO_2 \tag{8}$$

Eight chemicals are involved in the proposed reaction ($C_{diamond}$, $C_{graphite}$, C, $O_2$, $S_{liq}$, $SiC_{cubic}$, $SiO_2$, and CO). Local concentrations of these substances can be described by corresponding functions. The rates of change of these concentrations as functions of time at a specific point in the system can be expressed using a system of PDEs, the number of which is equal to the number of components.

$$\frac{\partial [C_{diamond}](x,t)}{\partial t} = D_{[C_{diamond}]} \frac{\partial^2 [C_{diamond}](x,t)}{\partial x^2} - ([C_{diamond}] * [O_2]) * k_I \tag{9}$$

$$\frac{\partial [C_{graphite}](x,t)}{\partial t} = D_{[C_{graphite}]} \frac{\partial^2 [C_{graphite}](x,t)}{\partial x^2} + ([C_{diamond}] * [O_2]) * k_I - ([C_{graphite}] * [Si]) * k_{III} + [CO]^2 * k_V \tag{10}$$

$$\frac{\partial [O_2](x,t)}{\partial t} = D_{[O_2]} \frac{\partial^2 [O_2](x,t)}{\partial x^2} - ([C_{diamond}] * [O_2]) * k_I + ([C_{diamond}] * [O_2]) * k_{II} - [O_2]^{1,5} * [SiC] * k_{IV} \tag{11}$$

$$\frac{\partial [Si](x,t)}{\partial t} = D_{[Si]} \frac{\partial^2 [Si](x,t)}{\partial x^2} - ([C_{graphite}] * [Si]) * k_{III} \tag{12}$$

$$\frac{\partial [SiC](x,t)}{\partial t} = D_{[SiC]} \frac{\partial^2 e(x,t)}{\partial x^2} + ([C_{graphite}] * [Si]) * k_{III} - ([SiC] * [O_2]^{1,5}) * k_{IV} \tag{13}$$

$$\frac{\partial [SiO_2](x,t)}{\partial t} = D_{[SiO_2]} \frac{\partial^2 [SiO_2](x,t)}{\partial x^2} + ([SiC] * [O_2]^{1,5}) * k_{IV} \tag{14}$$

$$\frac{\partial [CO_2](x,t)}{\partial t} = D_{[CO_2]} \frac{\partial^2 CO_2](x,t)}{\partial x^2} + ([CO]^2) * k_V \tag{15}$$

$$\frac{\partial [CO](x,t)}{\partial t} = D_{[CO]} \frac{\partial^2 [CO](x,t)}{\partial x^2} + ([O_2]^{1,5} * [SiC]) * k_{IV} - h^2 * k_V \tag{16}$$

A particular solution of the system of the ODEs is determined by the values of constants and initial conditions, in particular, by the initial arrangement of substances and their initial concentrations. Computational modeling was performed, and in the model system, substances were arranged as shown in **Fig. 3** (*t* = 0).

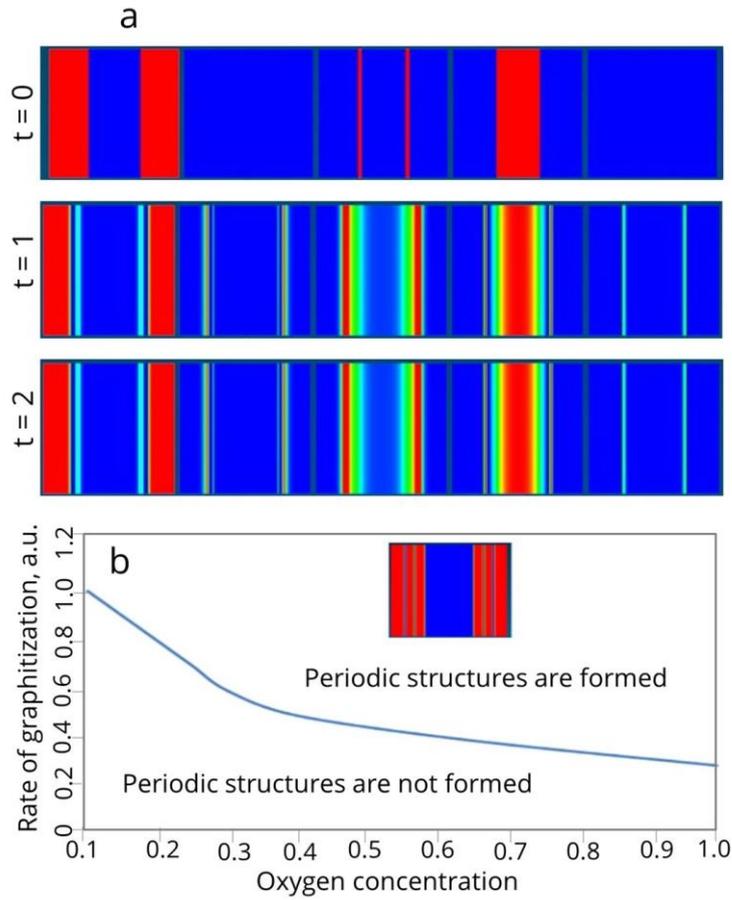

**Fig 3**. (a) A schematic of formation of silicon carbide on the diamond surface. The initial state corresponds to t = 0. (b) Influence of the graphitization rate and oxygen concentration on the formation of periodic structures.

At the first stage, the graphitization occurs (**Fig. 3a**, *t*=1). At the same time, the concentration of oxygen decreases, as it is spent on the formation of silicon and carbon oxides. Next, the graphite on the diamond surface reacts with silicon to form silicon carbide (**Fig. 3a**, *t* =1 and *t*=2). The influence of oxygen concentration and graphitization rate on the formation of periodic bands was investigated (**Fig. 3b**).

The rate of the graphitization is determined primarily by the temperature, and the graphitization is catalyzed by the presence of oxygen [20-23]. Both the simulation results and the observed experimental data show that a change in the crystallographic plane primarily changes the graphitization rate, therefore, the patterning process simply does not occur on some planes. On the other hand, as we see in **Fig. 1a**, patterns are formed only near the surface and there is no interaction of patterns formed by different diamond planes. Therefore, it will be reasonable to further simplify the model and analyze the flat (2D) case:

$$\frac{\partial u}{\partial t} = D_u \nabla^2 u - uv^2 + f(1-u) \qquad (17)$$

$$\frac{\partial v}{\partial t} = D_v \nabla^2 v + uv^2 - (f+k)v, \tag{18}$$

where $u$ and $v$ are the functions describing local concentrations of diamond and graphite, respectively, $D_u$, and $D_v$ are diffusion coefficients of diamond and graphite, $f$ and $k$ are constants, and $\nabla^2$ is the Laplace operator. Such a simplified model is used for various interfacial phenomena including friction-induced self-organization [24].

The first equation (Eq. 17) describes the process on the facet surface when the diamond turns into graphite at a rate of $f$. This process is autocatalytic, with gases absorbed on the diamond surface, metal impurities present in diamonds, as well as graphite nuclei serving as the catalyst for the graphitization process. The first equation (Eq. 18) describes the process of graphite removal from the system at the rate of $k$. During this process, graphite is converted to silicon carbide, which is considered a completely inert substance. The corresponding reactions are described by equation

$$D + nG \rightarrow mG \tag{19}$$

$$G \rightarrow SiC \tag{20}$$

As a result of the simulation of the system of Eqs. 17-18 for the 2D, the formation of patterns is observed, i.e. alternation of regions with different chemical compositions in the plane of the diamond–SiC interface, as shown in **Figs. 4a, b**. The Figures show a top view of the interface. If one cuts the resulting structure in half along the blue line and looks at the structure from the side, the view would be similar to the observed phase alternation in the **Fig. 4c**. Thus, we can speak of a qualitative agreement between the simulation results and the observed microstructures.

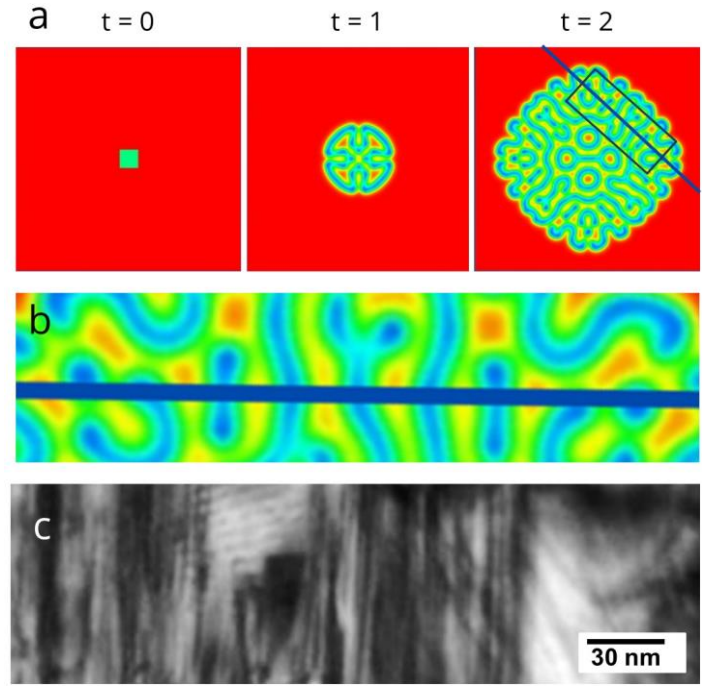

**Fig. 4**. (a) Formation of interconnected diamond–SiC structure by the reaction-diffusion processes (b) a higher resolution at *t* = 2 shows a pointed patterned structure. (c) TEM image showing an example of a similar patterned structure.

**Conclusions**

Modeling of the reaction-diffusion processes at the interface during the synthesis of diamond–SiC composite material allowed us to explain the formation of ceramic materials with a regular (periodic) interconnected microstructure, which was observed experimentally. Intercrystallite fractioning of the composite material with interconnected structure at the interface results in very high stiffness and superb mechanical properties including the resistance to impact loading. We have investigated both a simplified and a more detailed model describing the complex chemical reactions involved in the self-organization process. Both models are consistent with the formation of self-organized structures with the simplified model giving a qualitative picture. Our modeling approach can be generalized for other structures involving the Turing mechanism of self-organization.

**Funding:** This study was funded by the Russian Science Foundation (project No. 20-13-00054). TEM measurements were carried out on the equipment of the Core shared research facilities "Composition, structure, and properties of structural and functional materials" of the NRC «Kurchatov Institute» - CRISM "Prometey" with the financial support of the Ministry of Education and Science of Russia, agreement No. 13.CKP.21.0014 (unique identifier RF----2296.61321X0014).

**Conflict of Interest:** The authors declare that they have no conflict of interest.

**TOC Graphic**

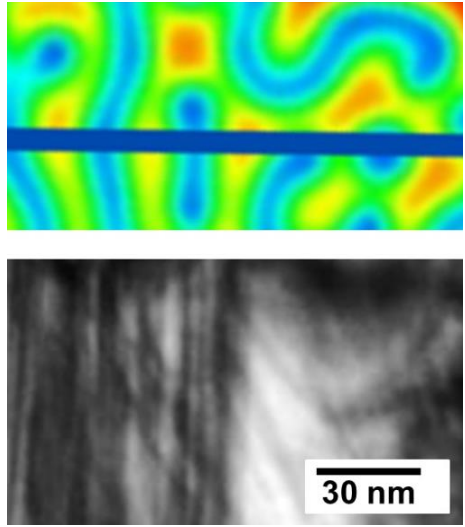